\documentstyle[12pt]{article}
\title{\bf Blue Horizontal--Branch Stars: The ``Jump'' in 
           Str\"omgren $u$, Low Gravities, and Radiative
           Levitation of Metals \footnote{Based on data collected at the 
        European Southern Observatory (La Silla, Chile) and 
        the Nordic Optical Telescope, Spain} }
\author{F.~Grundahl $^1$, M.~Catelan $^{2,3}$, W.~B.~Landsman $^4$,\\ 
        P.~B.~Stetson $^5$, and M.~I.~Andersen $^6$ 
\vspace{1cm}\\
\normalsize $^1$University of Victoria, Canada\\
\normalsize $^2$University of Virginia, USA\\
\normalsize $^3$Hubble Fellow\\
\normalsize $^4$Raytheon ITSS, NASA/Goddard Space Flight Center, USA\\
\normalsize $^5$Herzberg Institute of Astrophysics, Canada\\
\normalsize $^6$Nordic Optical Telescope, Spain }

\date{\mbox{}}
\begin{document}
\maketitle
\pagestyle{empty}
%
%
\def\bull{\vrule height .9ex width .8ex depth -.1ex}
\makeatletter
\def\ps@plain{\let\@mkboth\gobbletwo
\def\@oddhead{}\def\@oddfoot{\hfil\tiny\bull\quad
``The Galactic Halo: from Globular Clusters to Field Stars'';
35$^{\mbox{\rm rd}}$ Li\`ege\ Int.\ Astroph.\ Coll., 1999\quad\bull}%
\def\@evenhead{}\let\@evenfoot\@oddfoot}
\makeatother
%
%
\def\beginrefer{\section*{References}%
\begin{quotation}\mbox{}\par}
\def\refer#1\par{{\setlength{\parindent}{-\leftmargin}\indent#1\par}}
\def\endrefer{\end{quotation}}
%
%
{\noindent\small{\bf Abstract:} We study the ``jump'' in the blue 
horizontal--branch (BHB) distribution first detected by Grundahl
et al. (1998) in the Galactic globular cluster (GC) M13. On the 
basis of Str\"omgren photometry for a sample of fourteen GC's we
show that: 1) The jump is best characterized as a systematic shift, 
on a $u, u-y$ color-magnitude diagram, from canonical zero-age HB
(ZAHB) models, in the sense that the stars appear brighter and/or 
hotter than the models; 2) the jump is a ubiquitous phenomenon, 
ocurring over the temperature range 
11,500~K~$\leq {\rm T_{eff}} \leq$~20,000~K; 3) An analogous 
feature is present in ($\log\,g, \log\,{\rm T_{eff}}$)
diagrams -- indicating a common physical origin for the two 
phenomena; 4) The physical mechanism responsible for the jump
phenomenon is most likely radiative levitation of iron and other 
heavy elements.
 }
%
%
\section{Introduction}

It is of significant interest to understand what factors govern the 
morphology of the horizontal branch in stellar populations since it 
is often used as a distance and age indicator. In this contribution 
we will focus our attention on two problems in our understanding of
the HB that have recently become apparent: The occurence of BHB 
stars of lower gravities than predicted by canonical models (e.g., 
Moehler, Heber \& de Boer 1995) and the jump in Str\"omgren
$u$ first discovered in M13 by Grundahl, VandenBerg, \& Andersen 
(1998). 

Heber, Moehler, \& Reid (1997) suggested that helium mixing 
(Sweigart 1997a,b) in the HB progenitor stars might explain the 
low gravities problem; a similar suggestion was advanced by Grundahl  
et al. (1998) in regard to the $u$--jump phenomenon. Given that 
subsequent observations showed the $u$-jump to be present in all 
clusters with a sufficiently hot HB, we decided to undertake a 
systematic study of this phenomenon.  We shall demonstrate, in the 
following, that the low measured gravities and the $u$-jump are 
manifestations of one and the same physical mechanism which most 
likely is radiative levitation of heavy elements. The observations 
for this project and a more detailed account of the phenomenon  
are given in Grundahl et al. (1999). 

\section{The Ubiquitous Nature of the Jump}

In Fig. 1 we show the CMDs for all our observed clusters with 
a BHB. Several important conclusions can be drawn from 
this Figure: 1) the jump occurs in {\em all} clusters, irrespective
of any parameter characterizing these such as: metallicity, concentration, 
luminosity or extent of mixing on the RGB. This makes the
helium mixing hypothesis unlikely as the sole explanation for
this phenomenon; 2) the ``size'' of the jump is constant from 
one cluster to the next; 3) the onset of the jump occurs at 
${\rm T_{eff}} \approx 11,\!500\pm 500$~K for all clusters in 
the sample (derived from our calibrated photometry and Kurucz 
color-temperature relations); 3) The cool onset of the jump
appears very abrupt. Even in $\omega$ Centauri (which shows 
a significant metallicity spread) the jump is clearly present 
and well defined (as predicted in Grundahl et al. 1999); our 
new photometry for this cluster will be presented elsewhere. We 
also note that the jump occurs irrespective of the detailed 
HB morphology, i.e. whether the cluster has a long blue tail or
just a stubby BHB.

\begin{figure}
\vspace{20 cm}
\includegraphics{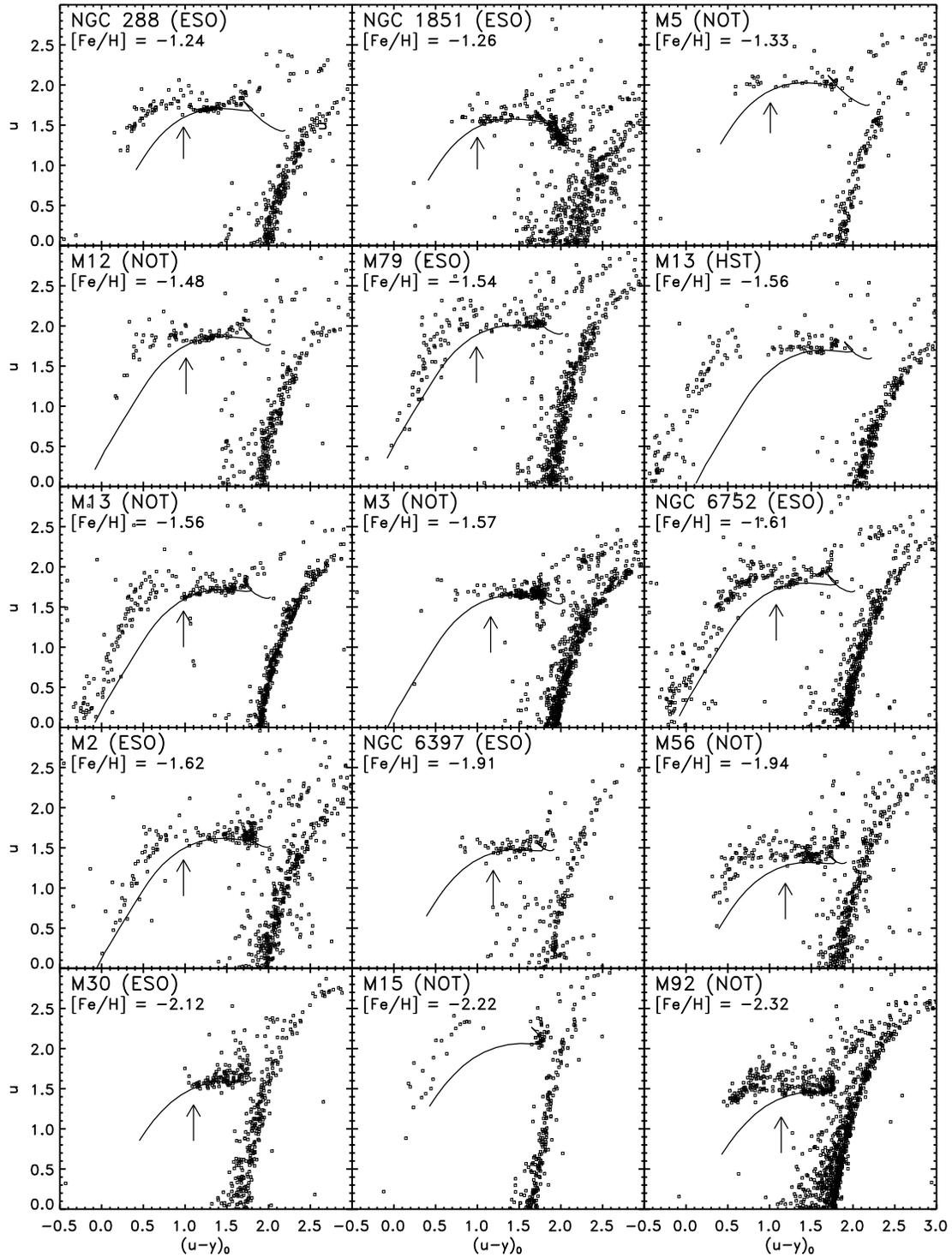}
\caption{ A mosaic plot showing all our CMDs. The vertical arrows indicate 
 the onset of the jump at its cool end.  }
\end{figure}

\section{Low Gravities and the Jump Phenomenon}

Several of our clusters have had stars with spectroscopically determined
gravities published in recent literature. Since our determination 
of the temperature for the onset of the jump appeared close to the 
region where gravities lower than predicted by theory were found we have 
investigated whether the two phenomena are related. Figure 2 shows
a plot of the stars in our sample with spectroscopic determinations
of their gravity. All stars which are classified as belonging to 
the jump region (based on the $uvby$ photometry) are plotted as 
black symbols, whereas stars located outside the jump region are
plotted as gray symbols. It is apparent from this Figure that stars 
classified as ``jump stars'' also have gravities lower than expected 
from ZAHB models. This clearly shows that {\em the two phenomena are 
related on a star by star basis} and hence that the two most likely 
are manifestations of the same physical phenomenon. As the effect 
seems unlikely to be caused by stellar evolution (no dependence on 
mixing history or age is found, both in terms of jump size and 
location) we strongly suspect that a stellar atmospheres effect 
is the cause. 

\begin{figure}
\vspace{10cm}
\includegraphics{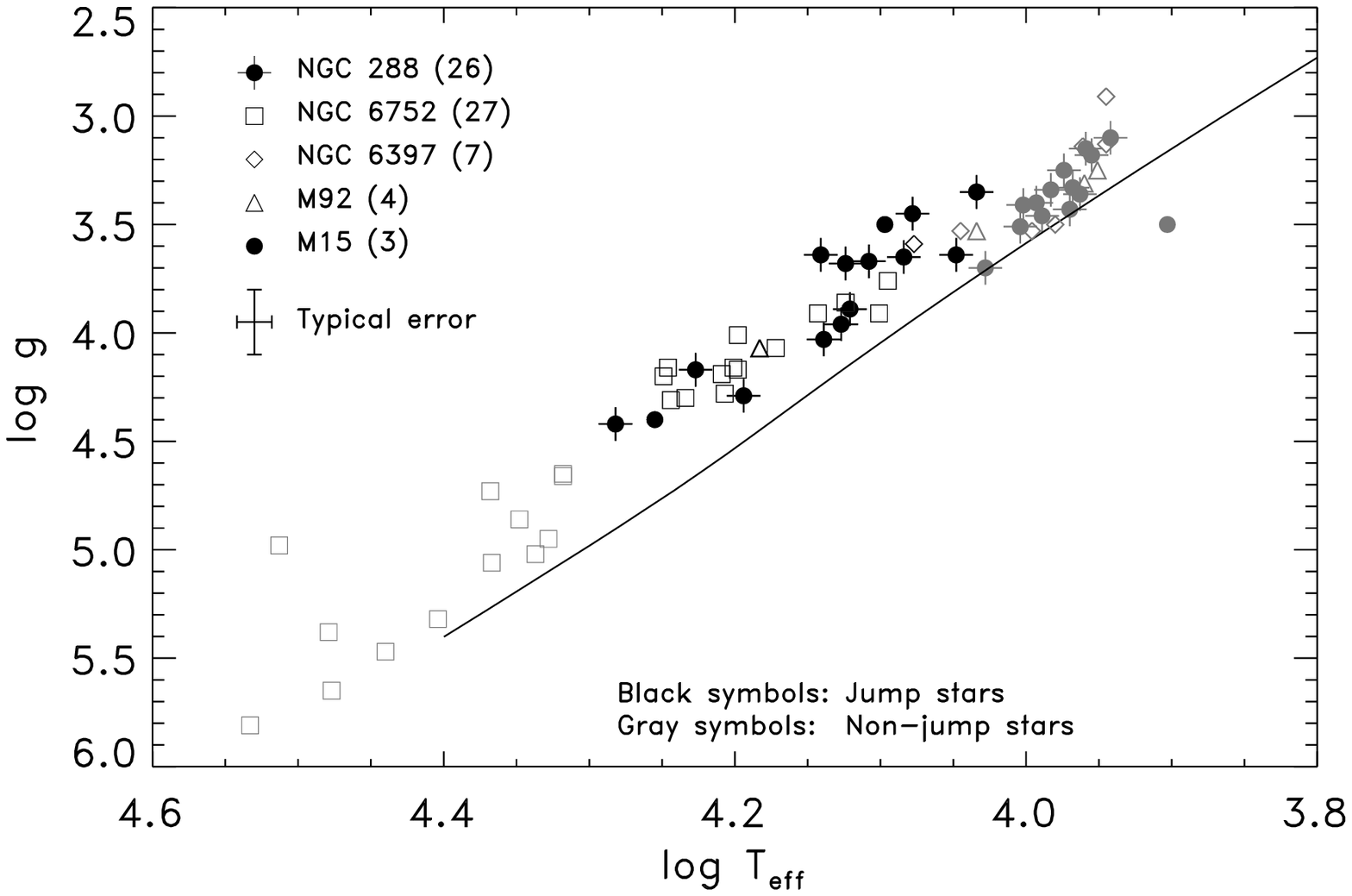}
\caption{
 Graphical demonstration of the close connection between the
 so-called low-gravity stars and the ``$u$-jump" stars. Stars 
 which lie inside the jump region in the CMDs of Figure 1 are
 plotted as black symbols, whereas stars which lie outside the 
 jump region are plotted with gray symbols.
        }
\end{figure}

\section{Radiative Levitation of Metals as an Explanation}

Glaspey et al. (1989) obtained high resolution spectroscopy of 
two BHB stars in NGC~6752 and found
that the one which lies inside the jump region at 16,000~K had
super--solar iron abundance whereas the one outside had normal
abundances compared to the other cluster stars. For field BHB
stars Bonifacio et al. (1995) and Hambly et al (1997) find 
similar trends with large over abundances of some of the heavy 
elements -- the detailed abundance patterns are quite complex. 

These results led Grundahl et al. (1999) to propose that radiative
levitation of heavy elements could be the cause for the $u$-jump
and low gravities phenomenon. Furthermore simple experiments with
enhanced heavy-element abundances, based on Kurucz solar-scaled
atmospheres (see Figure 3), succeeded in qualitatively producing
a higher flux in $u$ as seen observationally.  

\begin{figure}
\vspace{12cm}
\includegraphics{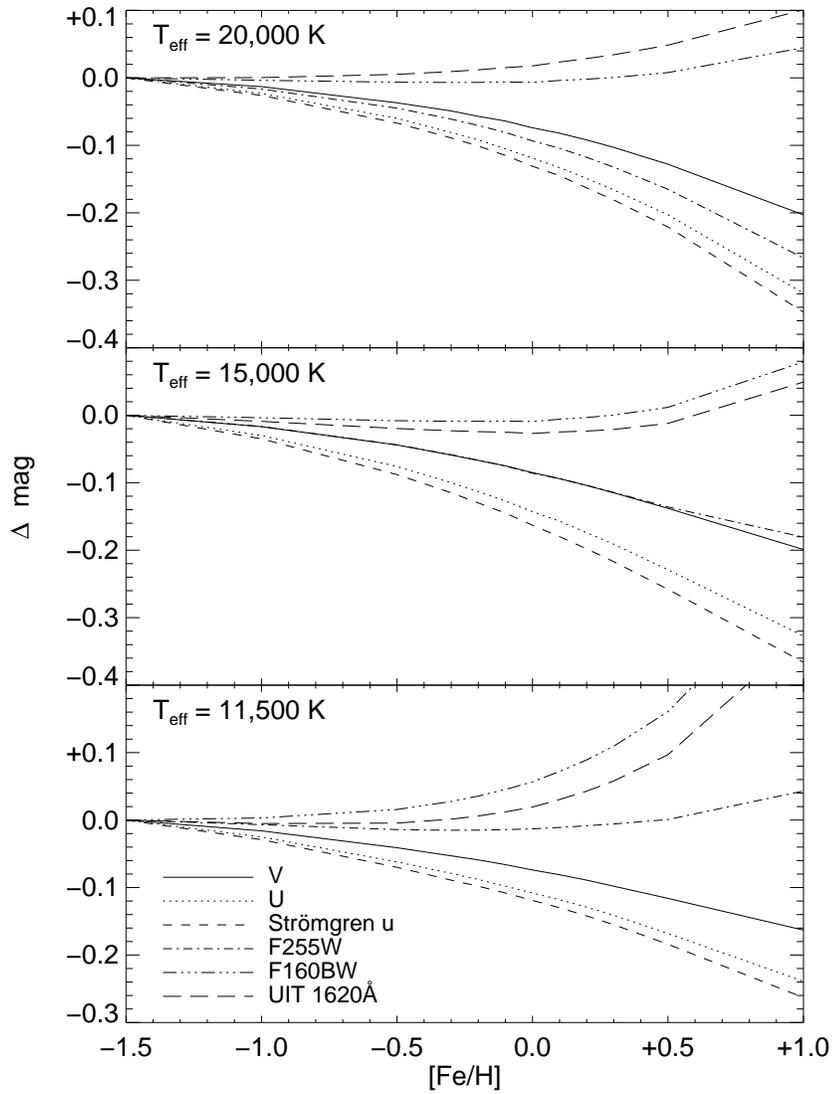}
\caption{
  The emergent flux in different bandpasses is shown as a 
  function of metallicity for Kurucz model atmospheres 
  with the indicated temperatures. Note that in all cases
  the largest effect of enhanced abundance occurs in the Str\"omgren 
  $u$ band, and that there is a marked temperature dependence 
  for filters in the mid-to-far UV (F255W, F160BW, and 
  UIT~1620~\AA). 
        }
\end{figure}

Subsequently this hypothesis was given strong observational 
support by the spectroscopic investigations of Moehler et al. 
(1999), Peterson (1999), and Behr et al. (1999). These studies 
showed that several of the heavy elements were overabundant by 
large factors in the atmospheres of BHB stars in NGC~6752 and 
M13. Behr et al. further found that the onset of the jump 
occurrs at ${\rm T_{eff}}\,\approx\,11,\!500$~K, in 
excellent agreement with our estimate. 
The study of hot HB stars in NGC 6752 by Moehler et al. (1999) 
further showed that the problem of too low gravities is 
significantly reduced (although not eliminated) if model 
atmospheres with appropriately high metallicity are used in 
the analysis of the spectra.

Given the remarkable similarity of the $u$ (and gravity) 
jump(s) from one cluster to the next, we suggest that 
very similar chemical abundance patterns as found by the 
above authors are likely present in the other systems as 
well. For detailed quantitative comparisons between observations
and theory in what concerns photometry and spectroscopy of blue 
HB stars lying inside the jump region, we warn that more 
realistic calculations taking into account the very complicated 
(but {\em observed}) abundance patterns -- using, e.g., Kurucz's 
ATLAS12 code -- will likely be necessary (Grundahl et al. 1999).

\section{Conclusions}

We have demonstrated that the $u$--jump discovered by Grundahl
et al. (1998) occurs in {\em every} globular cluster with HB stars
hotter than 11,500~K and that this phenomenon is connected, on
a star by star basis, to the ``low gravities'' found by 
Moehler et al. (1995). The most likely explanation 
is that radiative levitation of heavy elements into the 
stellar atmosphere changes the emergent flux pattern in 
such a way as to increase the $u$ flux and cause the measured
gravities to be too low if these element enhancements are not
taken into account. Detailed spectroscopic studies of large
samples of stars and theoretical diffusion calculations 
are urgently needed for further explanation of this phenomenon 
since ratiative levitation is expected to have implications 
for the interpretation of integrated ultraviolet spectra 
of old stellar populations -- both Galactic and extragalactic
(Landsman 1999).

%
%
\section*{Acknowledgements}
Support for M.C. was provided by NASA through Hubble Fellowship 
grant HF--01105.01--98A awarded by the Space Telescope Science 
Institute, which is operated by the Association of Universities 
for Research in Astronomy, Inc., for NASA under contract 
NAS~5--26555.
%
%
 
\beginrefer

\refer Behr B.B., Cohen J.G., McCarthy J.K., Djorgovski S.G., 
       1999, ApJ 517, L135

\refer Bonifacio P., Castelli F., Hack M., 1995, A\&AS 110, 441

\refer Glaspey J.W., Michaud G., Moffat A.F.J., Demers S., 1989, 
       ApJ 339, 926

\refer Grundahl F., VandenBerg D.A., Andersen M.I., 1998, ApJ 500, L179

\refer Grundahl F., Catelan M., Landsman W.B., Stetson P.B., 
       Andersen M.I., 1999, ApJ 524, in press (Oct.~10$^{\rm th}$ issue) 

\refer Hambly N.C., Rolleston W.R.J., Keenan F.P., Dufton P.L., 
       Saffer R.A., 1997, ApJS 111, 419

\refer Heber U., Moehler S., Reid I. N., 1997. In: Battrick B. (ed.) 
       ESA-SP 402, HIPPARCOS Venice'97, p.~461

\refer Landsman W. B., 1999. In: Hubeny I., Heap S., Cornett R. (eds.) 
       Spectrophotometric Dating of Stars and Galaxies. San Francisco,  
       ASP, in press (astro-ph/9906123)

\refer Moehler S., Heber U., de Boer K.S., 1995, A\&A 294, 65

\refer Moehler S., Sweigart A.V., Landsman W.B., Heber U., 
       Catelan M., 1999, A\&A 346, L1

\refer Peterson R. C., 1999, these proceedings 

\refer Sweigart A.V., 1997a, ApJ 474, L23

\refer Sweigart A.V., 1997b. In: Philip A.G.D., Liebert J., Saffer 
       R.A. (eds.) The Third Conference on Faint Blue Stars.  
       Schenectady, L. Davis Press, p.~3

\endrefer           
\end{document}